# Feature Engineering Approach to Building Load Prediction: A Case Study for Commercial Building Chiller Plant Optimization in Tropical Weather

Zhan Wang[a], Chen Weidong[a], Huang Zhifeng[a], Md Raisul Islam[a], Chua Kian Jon[a]

[a]Department of Mechanical Engineering, 21 Lower Kent Ridge Rd, 119077, Singapore

**Abstract**

In tropical countries where humidity levels are high, air-conditioning can account for up to 60% of a building's energy use. For commercial buildings with centralized air-conditioning systems, the efficiency of the systems largely depends on the performance of the chiller plant, and model predictive control is a promising strategy for optimizing chiller plant control due to its ability to make dynamic adjustments based on predictive load. The accuracy of the predictive load is undeniably important in such control strategies. The advantages of artificial neural networks in modeling nonlinear complex systems have made them a prominent data-driven method for load prediction. However, they are often prone to overfitting due to their complex layers and numerous parameters. Appropriate feature engineering can overcome this challenge. While weather data are key features for load prediction and have been used in almost all neural network models, they are often input as raw numerical values without being treated with advanced feature engineering methods. Clustering feature inputs is a promising feature engineering technique that can reduce model complexity while increasing prediction accuracy. Though previous studies have investigated using clustering algorithms on past cooling load and temperature for load prediction, no existing study has utilized them on multidimensional weather data. This highlights a research gap in applying clustering algorithms to transform raw weather data into new features that could improve prediction accuracy. Additionally, few studies have considered the uncertainties and noise in the cooling load profile. In response to these challenges, this study developed a cooling load prediction model that combines a neural network with Kalman filtering and K-means clustering. By applying the model to real-world data from a commercial skyscraper in Singapore's central business district, we improved the cooling load prediction accuracy by 46.5%. Furthermore, an optimal chiller sequencing strategy was developed for the studied building through genetic algorithm optimization based on the predictive load, leading to potential energy savings of 13.8%. Finally, the study also assessed the integration of thermal energy storage systems into the chiller plant design, demonstrating potential reductions in capital and operational costs by 26% and 13%, respectively.

*Keywords:* Cooling load prediction, Artificial neural network, Genetic algorithm, Thermal energy storage, Kalman filter, K-means clustering, Chiller optimization, Chiller sequencing

## 1. Introduction

The building sector is accounted for 39% of carbon emission globally and is the second largest energy consumption sector in Singapore. In alignment with the Paris Agreement, Singapore has committed to achieving net-zero emissions by 2050. To realize this commitment, the Singapore Green Building Master Plan has outlined ambitious targets, aiming to make 80% of buildings green and achieve an 80% improvement in energy efficiency for the highest standard green buildings by 2030. Inevitably, there exists substantial potential and pressing need for reducing energy consumption in the building sector.

Heating, ventilation, and air-conditioning (HVAC) systems play a significant role in the overall energy consumption of buildings. In the United States, for instance, these systems account for 50% of the energy consumption in buildings [1]. This figure can escalate to 60% in Singapore due to the prolonged need on air conditioning with the necessity of both sensible and latent heat removal in tropical climate [2]. Notably, chiller is the principal energy-consuming component within the air conditioning systems. A typical chiller plant is made up of chiller, cooling tower, pumps and its overall performance is commonly measured in kW/RT, which represents the electricity used by the plant for producing 1 refrigeration tonne of cooling effect to the building. The performance of chiller varies with the operating capacity, and centrifugal chillers are more efficient when its at full or nearly full load [3]. Hence, it is of best practices to operate chillers at the loading condition that consumes least energy. Optimization of the chiller plant aims to minimize the power consumption and achieve a lower kW/RT energy input while maintaining the needs and comfort of building occupants.

Traditionally, optimization of HVAC systems is done through supervisory control where techniques can be classified into model-free or model-based supervisory control. The commonly used model-free control strategies including expert systems and reinforcement learning face challenges of unstable behaviour due to incomplete knowledge database, intensive fine tuning and the dynamics of the HVAC systems. Model-based supervisory control, on the other hand, utilized tools to model the system component are responsive to changes in control parameters and enjoys the benefits of adapting to rapid changes in environment conditions [4]. Model predictive control(MPC) is a more advanced supervisory control strategy that emphasizes the influences of time varied environmental condition on system control effectiveness and incorporates future conditions in the present decision making and has demonstrated prominent performance in HVAC controls [5, 6, 7]. Intuitively, for implementing MPC on a chiller plant, the accurate prediction of building's cooling load is the first and crucial step.

Cooling load prediction methods can be broadly classified into physics-based and data-driven methods. The physics-based approach involves the creation of a detailed physics model based on fundamental heat transfer laws and comprehensive understanding of building characteristics. While this technique may capture the actual thermal response of the building, it encounters implementation challenges due to the considerable complexity of buildings and the lack of information updates over their operational life [8]. On the other hand, the data-driven method does not need to incorporate prior knowledge of the building characteristics and is built upon historical operational data collected from building's operational data [4]. The advantages of artificial neural network(ANN) in modeling non-linear complex systems and handling multi-variable problems have made it a prominent data-driven method for load prediction since 1990s [9, 10, 11]. In this study, we propose to incorporate feature engineering techniques that can potentially increase the prediction accuracy of ANN for building load prediction.

## 1.1. Previous Studies

As one of the pioneering works in using ANNs for load prediction, Kreider et al. [9] utilized data such as operating schedules, outside temperature, water temperature, food retailer sales, and a binary heating control signal to predict steam, water, electricity, and natural gas consumption. Since then, the ability of ANNs to provide accurate predictions without the need for building complex numerical simulations has made their adoption in load prediction common among researchers [12, 13]. Over the years, studies have explored the use of ANNs for predicting cooling load [8, 14, 15], heating load [16, 17], HVAC load [18, 19], electricity load [20, 21], and energy efficiency [22, 23]. Different ANN architectures have been investigated by previous studies including multilayer perceptron (MLP) [19, 24, 25], recurrent neural network(RNN)[26], long short-term memory(LSTM) [27], convolutional neural network (CNN)[28]. Besides building an effective model architecture, adopting feature engineering techniques is critical for developing an ANN model, as these transform raw data into meaningful features that ultimately improve the model's predictive performance.

Feature engineering is the process of selecting, transforming, and creating input features to improve the performance of machine learning models. Since neural networks are often prone to overfitting due to their complex layers and numerous parameters [29], feature engineering can reduce the model complexity while attaining the relevant information and patterns, hence helping to mitigate overfitting of ANN. Building cooling load can be broken down into three components: heat transferred through the envelope, heat produced by occupancy, and cooling and dehumidification of ventilation air. Existing studies have used weather conditions such as dry-bulb temperature and relative humidity as input features, due to their correlation with heat transferred through the envelope and ventilation air [30, 31], as well as day type and time features to account for occupancy load, given the daily seasonality in building load profiles [32, 33]. While some studies suggest that using a building's load at the previous time step can provide more reliable and accurate predictions due to the dynamics and auto-correlation of the building's load [34, 35], the ideal number of time steps often varies across studies, and there is no universal guideline for selecting the optimal number of past time steps. It is important to note that adding more features does not necessarily improve prediction reliability and accuracy, as too many features can lead to model overfitting and perform poorly on test dataset[26].

Clustering of feature inputs is a promising technique that can reduce model complexity while increasing prediction accuracy [47]. A clustering algorithm is a machine learning technique used to group similar data points into clusters, thereby uncovering inherent patterns in the data. When applied to ANNs, clustering can reduce dimensionality and highlight key features, leading to improved model performance [48, 49]. Previous studies investigating the application of clustering algorithms for building load prediction are summarized in Table 1, note that the applications also extend to energy classification, study of energy patterns, and occupant behavior, as these areas share similarities with load prediction. While most of the studies summarized used historical building load data with clustering algorithms, only a few investigated the use of outdoor temperature, and none explored clustering based on multiple weather parameters for building load prediction. Notably, the prediction model's ability in understanding and processing weather parameters plays an important role in short-term load forecasting [50]. And with the appropriate construction of these parameters, there is potential for developing adaptive control strategies that can response to the dynamics of building's load profile [51]. Hence, there exists a research gap in investigating the usefulness of clustering multi dimensional weather data for building load prediction, thereby unleashing the full potential of weather data for load prediction.

Table 1: Summary of studies on application of clustering algorithms for building load prediction

| Ref. | Features | Clustering Method | Application | Application Model |
|------|----------|-------------------|-------------|-------------------|
| [36] | AHU Operating parameters | K-means | MLP | short-term prediction of HVAC system |
| [37] | heating load, cooling load | K-means | MLP | predict heating and cooling energy efficiency |
| [38] | load, temperature, differenced load | K-means | Ensemble NN | building electricity load forecast |
| [39] | historical load | Fuzzy C-means | SVM | short-term load prediction of power systems |
| [40] | not specified | Fuzzy C-means | SVM | short-term cooling load prediction |
| [41] | building hourly consumption | K-shape | SVM | building energy prediction |
| [42] | building energy consumption | Entropy Weighted K-means | GESD | detect abnormal building energy consumption |
| [43] | building energy consumption | Fuzzy Clustering | N.A. | building energy classification |
| [44] | heat loss coefficient, temperature | K-means | N.A. | classify occupant behavior |
| [45] | building energy consumption | K-means | N.A. | identify rules for improving building performance |
| [46] | household energy consumption | K-means, K-medoid, SOM | N.A. | study household characteristics with energy use patterns |

Table 2 presents a review of more recent studies that have enhanced neural network predictions through feature engineering. Although weather data have been used in all studied models, they are typically treated as raw numerical values, and the potential of using machine learning algorithms to transform them into more meaningful features has not been fully explored. In this study, we propose to adopt K-means clustering to transform weather data into new features for load prediction. Despite the fact that some previous studies have utilized K-means clustering for feature engineering [52, 53, 54, 55], the specific impact of clustering weather data on neural network prediction accuracy remains under-investigated. Additionally, the noise resulting from the dynamics of chiller operation and inherent measurement errors remains an ongoing challenge in load prediction [56, 57]. To address this, we propose implementing a Kalman filter to mitigate the inherent noise in cooling load measurements, further improving the accuracy of load prediction.

*1.2. Proposed Study*

Figure 1 illustrates the overall workflow of this study. First, a month of chiller plant's operation data from a 65-storey mixed-use commercial skyscraper located at Singapore central business district is sampled for this study. Second, the collected data are processed and prepared into feature sets incorporating feature engineering techniques including Kalman Filter and K-means clustering. Third, the training and validation of the neural network models is developed and compared between two commonly used model architecture, namely multi-layer perceptron (MLP) and long short-term memory (LSTM), following which the best performing validation model is used for load prediction on test dataset. A new chiller sequencing strategy targeted at minimizing chiller energy consumption is then derived from a optimization model built upon chiller partial load efficiency profile using genetic algorithm. Finally, the energy and cost benefits of integrating a thermal storage system into the chiller plant design is evaluated as a holistic approach to reduce the carbon footprints of the commercial building.

The rest of this study is structured into three sections, section 2 explains the methodology, where 2.1 elaborates on the data source, 2.2 explains on the feature engineering techniques, 2.3 details on structure of the neural network models, and 2.4 illustrates on the formalization of the optimization problem. Section 3 presents and discusses the results from the prediction model (3.1), optimization model (3.2, and integration with thermal energy storage (3.3). Section 4 concludes and highlights the main contributions of this study.

Table 2: Recent studies that have enhanced neural network predictions through feature engineering

| Ref. | Model | Features | Findings |
|---|---|---|---|
| [58] | MLP | date and time; past outdoor temperature; past cooling load | feature engineering with big data analysis can reduce errors in cooling load demand prediction |
| [59] | LSTM; MLP | hour; outdoor temperature; humidity; past cooling load | LSTM is useful for feature extraction; Past cooling load has the greatest impact for load prediction |
| [60] | MLP | past cooling load; outdoor temperature; humidity; occupancy level | occupancy level can modeled based on electricity consumption |
| [8] | DNN | past cooling load; outdoor temperature; humidity | deep autoencoder for feature extraction boosts performance for nonlinear models |
| [61] | LSTM | historical occupancy data; building electricity usage; date | longer input time lags cause instability in short-term prediction |
| [62] | LSTM | hour; day; holiday; outdoor temperature; humidity | LSTM is more robust to input data uncertainty |
| [63] | CNN; LSTM | historical energy consumption; weather parameters | DNN enhanced feature selection improves prediction accuracy |
| [64] | LSTM; MLP | time; weather parameters; HVAC system parameters; electricity price | features extracted by PCA can improve model performance |
| [65] | MLP | time; day; weather parameters; historical occupant data | SARIMA can capture temporal sequential feature of the data |
| [66] | MLP; SVR; XGB | historical cooling load; weather parameters; time | exhaustive search of optimal input feature combination selects the best input features |

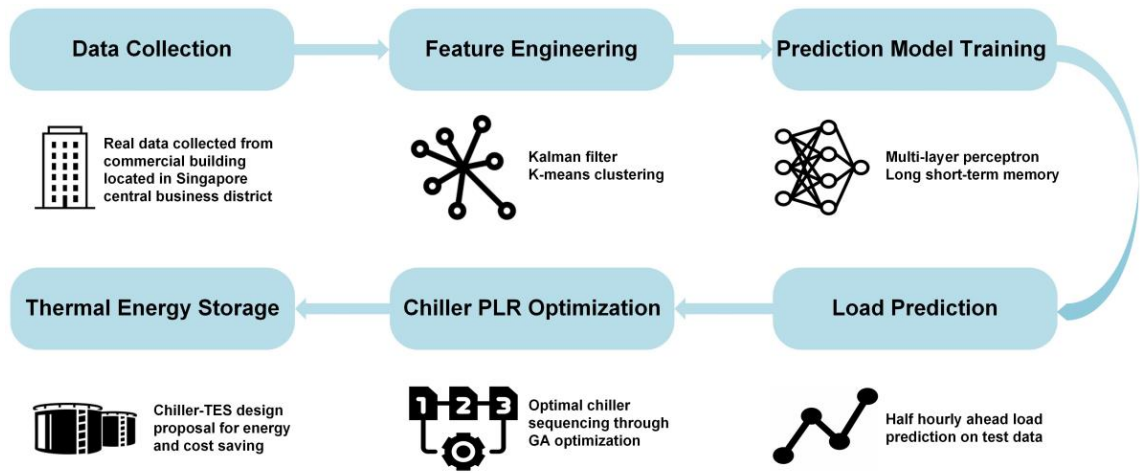

Figure 1: Overall workflow of the study

## 2. Methodology

### 2.1. Data Collection

A commercial mixed development located in the central hub of Singapore is sampled for purpose of the study. The development comprises four commercial uses: office, retail, residential, and hotel. The office strata has a total gross floor area of nearly 100,000 square meters, while the retail strata covers approximately 15,000 square meters. Both the office and retail strata share a common chiller plant, while the hotel and residential strata have individual air-conditioning systems. This study specifically utilizes data from the combined office and retail strata, collectively referred to as "the building" throughout the remainder of the study.

The air-conditioning system of the building is supported by a water-cooled chiller plant with a rated capacity of 5,000 refrigeration tons(RT). Six variable speed centrifugal chillers are deployed onsite, four of which are of 1000 RT capacity and the others are of 500 RT capacity each. The chilled water loop, which circulates chilled water to the air-handling units in the office strata and fan coil units in the retail strata, is facilitated by six chilled water pumps. Additionally, six condenser water pumps transport condenser water to the cooling tower, where the heat is expelled into the atmosphere. The building operates 24 hours a day, year-round, as chilled water is required for the server rooms of office tenants during nighttime hours. The standard operational hours are from 9 am to 6 pm for the office strata and from 10 am to 9 pm for the retail strata.

One months' operational data from the chiller plant is collected from the building management system, spanning from 1 August 2023 till 31 August 2023 at one minute interval. The data collected includes chilled water supply temperature, chilled water return temperature, chilled water header flow rate, condenser water supply temperature, condenser water return temperature, condenser water return header flow rate, power consumption in kW of the respective water pumps and cooling towers. The cooling load of the building is calculated using the formula *CoolingLoad* = $C_{p,w} \cdot m_w \cdot (T_R - T_S)$, where $C_{p,w}$ is the specific heat capacity of water, $m_w$ is the mass flow rate of the chilled water, $T_R$ is the chilled water return temperature and $T_S$ is the chilled water supply temperature. Weather data were extracted from the database the nearest weather observation station at Changi at half-hourly intervals. The data includes temperature, humidity, wind direction, wind speed, and pressure.

*2.2. Feature Engineering*

Selecting appropriate feature sets is crucial in prediction models because they: (a) improve prediction accuracy, (b) enhance the computational efficiency of training models, and (c) offer insights into data generation mechanisms. The cooling load of a building is typically from three components, heat transferred through the envelope, heat produced by occupancy, cooling and dehumidification of ventilation air. Heat transferred through roof and walls are driven by the temperature differences and the material's thermal resistance by the Fourier's Law. Since the properties of the building envelope remain unchanged and the indoor temperature is typically maintained at a constant level, the primary driving factor for heat transfer through the roof and walls is the outdoor temperature. Additionally, wind speed affects the convection heat transfer coefficient of the building envelope and is incorporated as a feature in the study. Heat gain through windows is influenced by the orientation of the building, the sun's angle, and atmospheric conditions such as cloud coverage. While the building's orientation is fixed and the sun's angle depends on time, cloud coverage is uncontrollable and challenging to quantify. Heat generated by occupancy fluctuates depending on the nature of human activities within the building and is contingent on building occupancy. Given the unavailability of specific occupancy data, the occupancy profile for a commercial building with office and retail functions is typically set for weekdays and weekends, correlated with time. Day type and time are utilized as variables to account for the influence on the human, lights, and equipment load. Cooling and dehumidification of ventilation air involves both sensible and latent heat gain. Sensible heat gain is driven by temperature differences, whereas latent heat gain is driven by the humidity ratio difference. In tropical countries like Singapore, where maintaining indoor humidity within a comfortable range requires the removal of more moisture, latent heat gain often constitutes a substantial portion of the total cooling load.

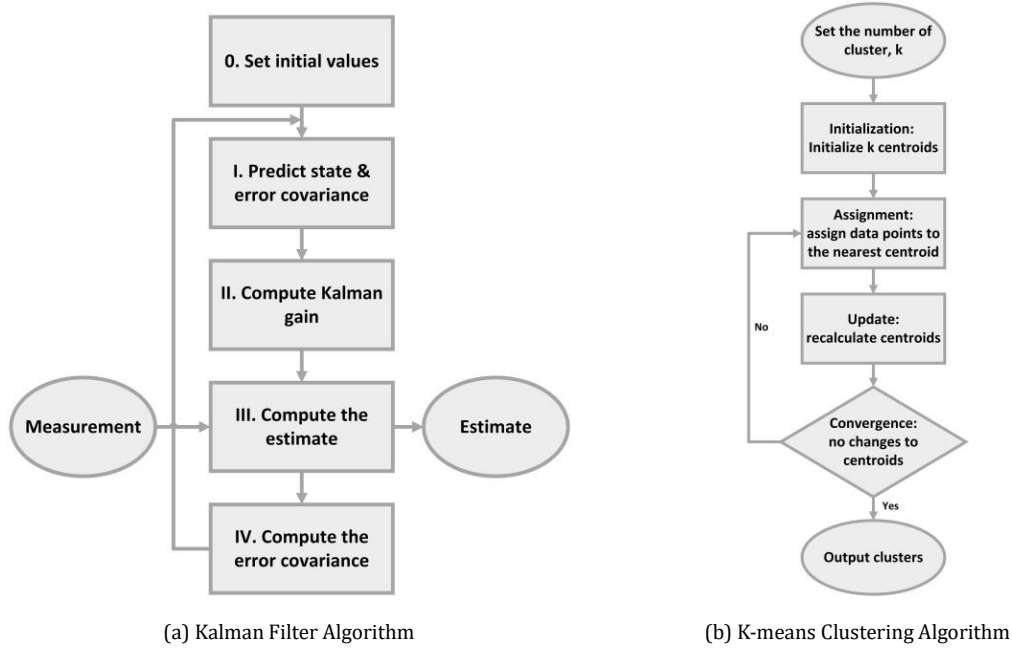

(a) Kalman Filter Algorithm     (b) K-means Clustering Algorithm

Figure 2: Kalman Filter and K-means Clustering Algorithm

In the context of cooling load prediction, feature selection faces several challenges: (a) balancing the model's complexity and the number of features to avoid overfitting or underfitting. Overfitting leads to high training accuracy but low testing accuracy, whereas underfitting fails in modeling complex systems accurately; (b) the varying nature of buildings means that the same features can have different impacts across different buildings; (c) potential noise introduced by the measurements from building automation systems [67]. In this study, two novel methods is proposed to address the challenges. The noise in measurements is managed by Kalman Filter and the complexity of the prediction model is brought down by introducing unsupervised machine learning algorithm K-means Clustering on weather data.

*2.2.1. Kalman Filter*

Chiller cut in may results in heavy fluctuations in building's cooling load profile. This phenomenon arises when a new chiller initiates operation, initially producing chilled water at a temperature higher than the desired setpoint. Consequently, this leads to an abrupt increase in the chilled water supply temperature and a sharp decrease in the calculated cooling load. These fluctuations pose challenges for load prediction. To address this, a common statistical method, the Kalman Filter, is employed.

The Kalman Filter is adept at producing the best estimate of the system's true current state and is initially applied to the raw cooling load dataset. The Kalman filter algorithm is illustrated in Figure 2a and it can be broke down into a prediction stage and an estimate stage. Step *I* represents the prediction stage, the state prediction and error covariance are calculated by

$$\hat{x}_k = A\hat{x}_{k-1}$$
$$P_k^- = AP_{k-1}A^T + Q$$

where *A* is the state transition matrix, $P_k$ denotes the predicted covariance of the state estimate error, providing a measure of the estimated accuracy of the state prediction and *Q* is the process noise covariance matrix, representing the uncertainty of the system dynamics. The estimate stage involves step *II* and *III*. In step *II*, the Kalman gain is computed by

$$K_k = P_k^- H^T (HP_k^- H^T + R)^{-1}$$

where the Kalman gain $K_k$ is a factor that balances the weight given to the predicted state and the new measurement, H is the measurement matrix that relates the state to the measurement, and R is the measurement noise covariance matrix, indicating the uncertainty in the measurements. In step *III*, the estimate is computed by

$$\hat{x}_k = \hat{x}_k^- + K_k(z_k - H\hat{x}_k^-)$$

where $z_k$ is the measurement at time $t_k$. In step *IV*, the error covariance gets updates by

$$P_k = (I - K_k H)P_k^-$$

in every loop. The state transition matrix A is set to 1 as it is assumed that the closest estimate of the current prediction is the last estimate. As the cooling load is directly measurable, the measurement matrix H is set to 1. Q and R matrices are also set to 1 assuming there is no process noise and measurement noise.

*2.2.2. K-means Clustering*

Weather information is often used in prediction model as the dynamic changes in building cooling load is often related to temperature, humidity [60, 8]. However, the conventional way of inputting weather data as individual features and in numerical format has a few challenges: a) as there are more features input to the system, it unnecessarily complex the model. This will lead to more computation time that is not desirable. b) inputting weather data in raw format may not yield the best test performance. Clustered weather data as input to the model is proposed in this study to address these challenges.

K-means clustering is an unsupervised machine learning technique that groups objects with more similarities into the same cluster and the algorithm is visualized in Figure 2b. In the Initialization step, $k$ random points are selected as cluster centers, called centroids. These centroids are denoted as $\mu_1,\mu_2,...,\mu_k$. Next in the Assignment step, each observation point is assigned to the cluster with the nearest centroid. This is done by calculating the distance between each observation and each centroid. The commonly used distance metric is the Euclidean distance. The assignment of an observation $x_i$ to cluster $j$ is represented as:

$$\text{Cluster}(x_i) = \arg\min_{j} \|x_i - \mu_j\|^2$$

The update step recalculates the centroids of the clusters by taking the average of all points assigned to that centroid's cluster. The update of centroid $\mu_j$ is calculated as:

$$\mu_j = \frac{1}{|S_j|} \sum_{x_i \in S_j} x_i$$

where $S_j$ is the set of all points assigned to the $j$-th cluster. The assignment and update steps will be repeated until the centroids no longer change, indicating that the clusters are as stable as possible and the algorithm has converged.

The effect of K-means clustering on the prediction model is evaluated on ambient dry bulb temperature, humidity ratio, and wind speed and compared with normalized raw weather data .

*2.2.3. Construction of Feature Sets*

Figure 3 summarizes the construction of feature sets. Eight sets of features are created as model inputs, namely: *Raw N = 1*, *Raw _N = 5*, *K = 2 _N = 1*, *K = 2 N = 5*, *K = 3 _ N = 1*, *K = 3 N = 5*, *K = 4 _N = 1*, *K = 4 _N = 5*, where Raw represents feature sets with raw weather data being z-score normalized, *K* refers to the numbers of clusters used in K-means clustering, *N* refers to the historical cooling load in *N* time steps (includes the current state, e.g., *N* = 1 refers to the current time step, N=5 refers to the current time step and the past 4 time steps). Using historical cooling load data to forecast future cooling load is a commonly used technique for load prediction [68, 67]; however, the ideal number of time steps may vary between models and datasets. Therefore, the suitability of this technique is being investigated by comparing $N = 1$ and $N = 5$ in the proposed model. Other input features include workday/off-day encoded in a one-hot categorical format and normalized time data. Due to the limited time interval of the available weather data, the output target is set as the half-hourly ahead cooling load.

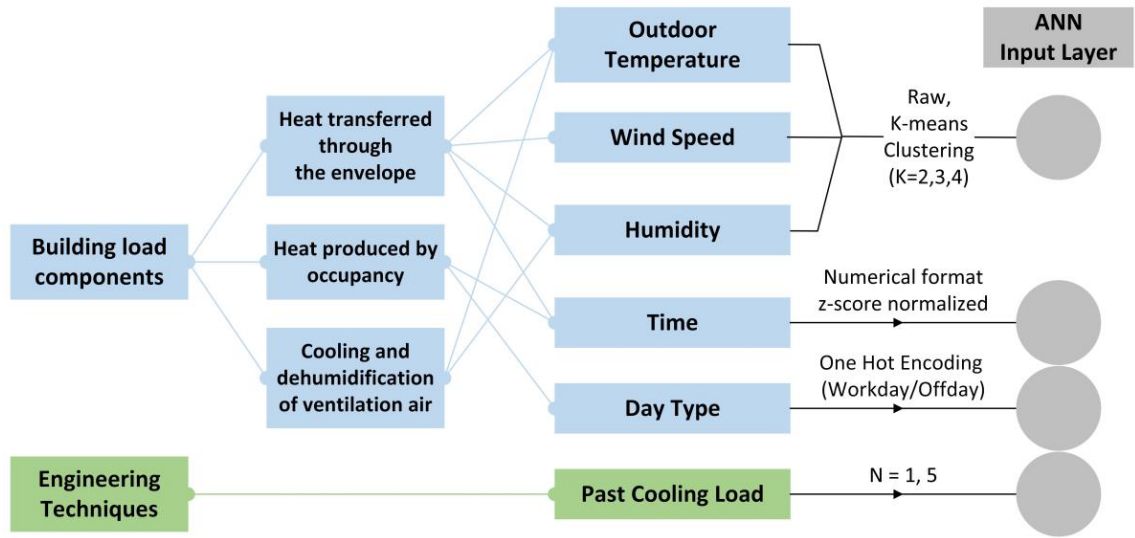

Figure 3: Construction of Feature Sets

A benchmark feature set is established to assess the performance of the 8 feature sets. The benchmark is constructed with cooling load without Kalman Filter, weather data without K-means clustering, and only the current state cooling load to predict the half-hourly ahead cooling load.

*2.3. Prediction Model*

*2.3.1. Multi-layer Perception*

A feedforward MLP network with one hidden layer is constructed in MATLAB using the feedforwardnet function on an 11th Gen Intel(R) Core(TM) i7-1165G7 machine with 16GB RAM. Previous studies have shown that increasing the number of layers does not significantly improve performance but does increase computational demand [69]. The network uses Levenberg-Marquardt backpropagation as the training function, and reasonable hyperparameter selections were made through manual tuning for the sake of brevity. However, it is acknowledged that auto-tuning could potentially enhance the model's performance.

The training process of MLP is illustrated in Figure 4. A feedforward network processes information stored in the input layer, containing input features. This information is then passed in one direction to the hidden layer and subsequently to the output layer, with different weights connecting them. Neurons in the hidden layer incorporate non-linear activation functions, introducing non-linearity to the model.

After being processed by the activation functions in the hidden layer, signals are passed to the output layer with another set of weights connecting the two layers. The errors between the predicted output and the actual target is calculated by the loss function

$$J = \frac{1}{m}\sum_{i=1}^{m}(y_i - \hat{y}_i)^2$$

where *m* is the number of training examples, $y_i$ and $\hat{y}_i$ are the actual and predicted values. The errors are then processed through the optimizer to update the weights, known as backpropagation. The fundamental formula governing the weight updates in the backpropagation algorithm can be expressed as

$$W_{ij}^{(l)} = W_{ij}^{(l)} - \alpha \frac{\partial J}{\partial W_{ij}^{(l)}}$$

where $W_{ij}^{(l)}$ represents the weight between neuron *i* in layer *l* and neuron *j* in layer *l* + 1, *ff* is the learning rate, and $\frac{\partial J}{\partial W_{ij}^{(l)}}$ denotes the partial derivative of the cost function *J* with respect to the weight. The training of the weights iterates until the loss function is minimized or the maximum training epochs is achieved.

*2.3.2. Long Short-term Memory*

LSTM is a type of Recurrent Neural Network (RNN) designed to capture underlying sequential relationships in time series data by preserving information processed in past time steps, achieved through a specialized gating system consisting of an input gate, an output gate, a forget gate as illustrated in Figure 5. These gates collectively help the network to retain or forget information over long periods, addressing the vanishing gradient problem found in standard RNN.

The forget gate layer decides which information is to be retained and which discarded. For a given LSTM unit at time step *t*, the forget gate $f_t$ is updated using the equation:

$$f_t = \sigma(W_f \cdot [h_{t-1}, x_t] + b_f)$$

where denotes the sigmoid function, $W_f$ are the weights applied to the forget gate, $h_{t-1}$ is the hidden state from the previous time step, $x_t$ is the input vector at the current time step, and $b_f$ represents the bias term for the forget gate.

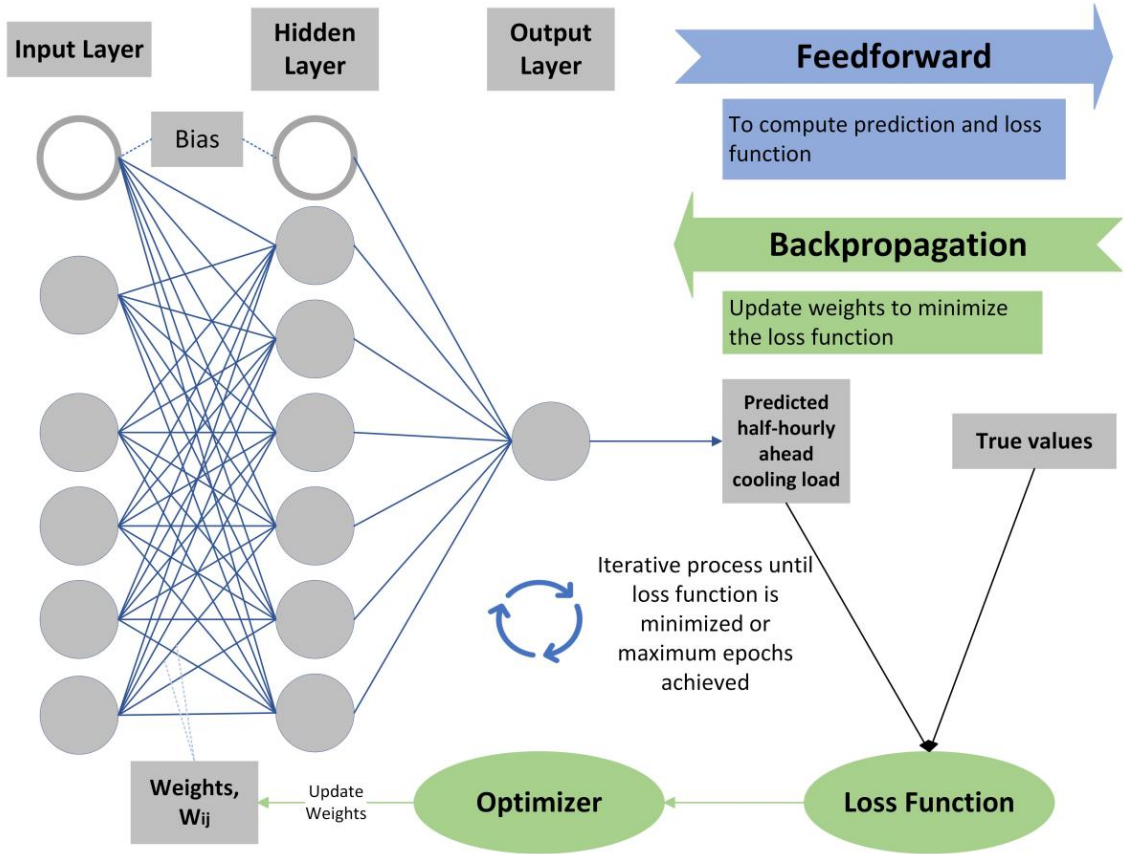

Figure 4: Training Process of MLP

The input gate layer decides which values will be updated with new information from the current input. The input gate $i_t$ and the candidate values $\tilde{C}_t$ are updated as follows:

$$i_t = \sigma(W_i \cdot [h_{t-1}, x_t] + b_i)$$
$$\tilde{C}_t = \tanh(W_C \cdot [h_{t-1}, x_t] + b_C)$$

where $W_i$ are the weights applied to the input gate, $W_C$ are the weights applied to create candidate values for the cell state, $b_i$ and $b_C$ represent the bias terms for the input gate and cell state candidate respectively.

The cell state $C_t$ is updated by forgetting the information deemed unnecessary and adding new candidate values scaled by the input gate:

$$C_t = f_t * C_{t-1} + i_t * \tilde{C}_t$$

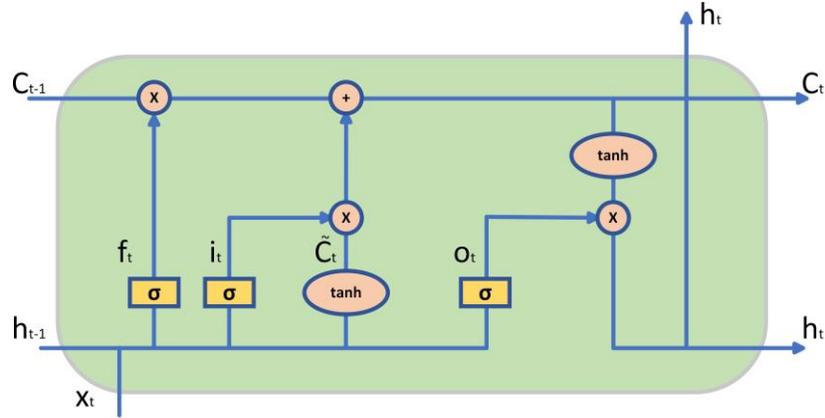

Figure 5: Gating system in LSTM

where * denotes element-wise multiplication.

The output gate layer decides what the next hidden state should be. The output gate $o_t$ and the hidden state $h_t$ are updated as follows:

$$o_t = \sigma(W_o \cdot [h_{t-1}, x_t] + b_o)$$

$$h_t = o_t * \tanh(C_t)$$

where $W_o$ are the weights applied to the output gate and $b_o$ represents the bias term for the output gate.

*2.4. Optimization Problem*

The goal of optimizing the chiller plant is to achieve minimum energy consumption while meeting the cooling load demand of the building. For development with multiple chillers, conventional chiller sequencing strategy is usually dependent on the chilled water supply temperature and flow rate. For example, for a primary pumping system where there's only one set of chiller water pump to support the chilled water loop, additional chiller will cut in when chilled water supply temperature exceeds the set point, or the measured cooling load meets or exceeds the operating chiller capacity. This conventional strategy ensures that the cooling load for the building can always be supplied by the chiller plant without considering the efficiency aspect, the reason is due to that chillers have different efficiency varying with part load ratio (PLR), where PLR of the chiller is defined as:

$$\text{PLR}_i = \frac{\text{chiller operation load}_i}{\text{chiller capacity}_i}$$

Chiller power can be represented by the polynomial functions of the part load ratio (PLR) [70, 71, 72]:

$$P_i = a_i + b_i \cdot \text{PLR}_i + c_i \cdot \text{PLR}_i^2 + d_i \cdot \text{PLR}_i^3$$

where $a_i$, $b_i$, $c_i$, and $d_i$ are interpolation coefficients relating the consumed power to the PLR for the $i$th chiller.

As chiller's operating efficiency is dependent on chiller loading, and it varies with brand, type of compressor and refrigerant used, it is hence important that chiller sequencing strategy can ensure chillers are operated in the manner that the minimum energy consumption can be achieved while ensuring that the cooling load of building is matched. Mathmatically, the optimization problem can be expressed as:

$$P_{total} = \sum_{i=1}^{k} P_i(PLR_i)$$

where $P_{total}$ is the total power, $P_i$ is the power of chiller $i$, and $PLR_i$ is the part load ratio of chiller $i$. The optimization problem is subject to the constraint that the building cooling load must be met by total chiller operation load:

$$\sum_{i=1}^{k} \text{PLR}_i \cdot \text{chiller capacity}_i \geq \text{Cooling Load}$$

Considering the suggestions of the manufacturer, the PLR of the chiller should not operate below 0.3. Hence, the optimization problem is subject to the second constraint:

$$0.3 \leq \text{PLR}_i \leq 1 \quad \text{or} \quad \text{PLR}_i = 0$$

*2.4.1. Genetic Algorithm*

The GA is an evolutionary algorithm inspired by the process of natural selection and the algorithm is as illustrated in 6. Initially, the algorithm generates a diverse population of PLR settings randomly. Each set in this population acts as a potential solution, represented as a vector of PLR values for the chillers. The algorithm evaluates the fitness of each individual by using an objective function, aimed at minimizing the total energy consumption.

Selection for reproduction is based on fitness, where individuals with higher fitness are more likely to be chosen, facilitating the spread of desirable characteristics. Through crossover, the genetic information of two parents merges to create offspring, whereas mutation introduces random changes, fostering new explorations in the search space.

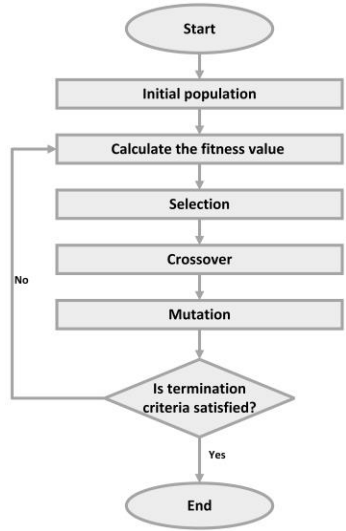

Figure 6: Genetic Algorithm

The algorithm iterates through its cycle until it either achieves a predetermined maximum number of generations or attains an acceptable fitness level. This level of fitness acceptability is determined when the average change in the fitness value drops below the predefined function tolerance, or when the constraint violation diminishes to less than the established constraint tolerance.

## 3. Results and Discussion

### 3.1. Load Prediction

#### 3.1.1. Prediction Performance

The feature sets are trained with the methodology illustrated in section 2.3. Each feature set is divided into a training set, validation set, and test set with a ratio of 70%, 15%, 15%. The model with the best validation performance from 10 runs on each feature set is then applied to the test set for prediction performance evaluation. Prediction performance is evaluated using the root mean squared error (RMSE) metric defined as

$$RMSE = \sqrt{\frac{\sum_{i=1}^{n}(y_i - \hat{y}_i)^2}{n}}$$

where $y_i$ and $\hat{y}_i$ are the actual and predicted values at time *i* respectively, and *n* is the total observation number.

Table 3 summarizes the resulting prediction performance of the benchmark feature set and the 8 feature sets on MLP and LSTM models. Overall, the performance of the $K = 2$ _$N = 1$ feature set with a combination of Kalman Filter and K-means clustering achieves a 46.5% improvement against the benchmark on the MLP model and 41.8% on the LSTM model.

Table 3: Summary of Prediction Performance on Test Dataset (RMSE)

| Model | Benchmark | Raw-N=1 | Raw N=5 | K=2 N=1 | K=2 -N=5 | K=3 N=1 | K=3 -N=5 | K=4 -N=1 | K=4 -N=5 |
|---|---|---|---|---|---|---|---|---|---|
| MLP | 153.6247 | 118.1734 | 93.9638 | **82.1567** | 91.3361 | 88.8217 | 120.4315 | 91.366 | 112.6679 |
| LSTM | 157.564 | 95.6193 | 104.4386 | **91.742** | 100.3647 | 98.6305 | 106.6211 | 102.7794 | 95.4689 |

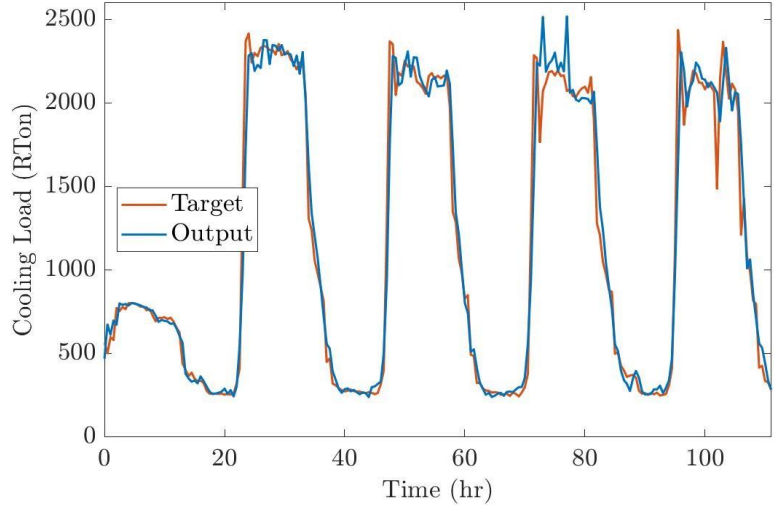

Figure 7: Benchmark MLP Model Test Performance

Figure 7 and 8 presents the comparison of benchmark MLP model and the $K = 2$ $N = 1$ MLP model. The performance comparison demonstrates that incorporating Kalman filtering and K-means clustering provides clear improvements in cooling load prediction accuracy. The benchmark MLP model, though capable of capturing the overall cooling load pattern, exhibits notable deviations from the target values, especially during periods of peak and valley load. In contrast, the $K = 2$ _$N = 1$ MLP model, which utilizes Kalman Filter to smooth the noisy raw cooling load and K-means clustering to better structure the feature set, significantly reduces the error range and improves the alignment of the predicted values with the actual cooling load. The prediction curve of this model more closely follows the target, with fewer and smaller deviations. This is demonstrated in the RMSE for the $K = 2$ _$N = 1$ model is markedly lower at 82.1567 RTon, nearly half that of the benchmark MLP model.

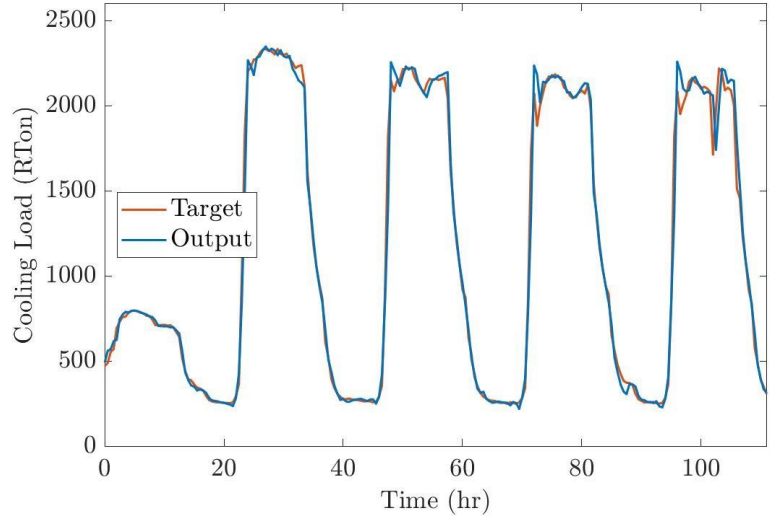

Figure 8: $K$ = 2 $N$ = 1 MLP Model Test Performance

### 3.1.2. Effect of Kalman Filter

The application of the Kalman Filter effectively smooths the fluctuations observed in the raw cooling load profile, particularly during the chiller cut-in periods. As shown in Figure 9, the fluctuations present in the cooling load measurements are smoothed, resulting in a more stable and accurate representation of the cooling load.

Kalman Filter managed the noise presented in the cooling load measurement and improved the prediction performance. Comparing the performance of *Raw N* = 1 feature set against the benchmark on both MLP and LSTM models, where the difference between the two feature sets is only the use of cooling load measurement and Kalman Filter cooling load, it is demonstrated 23.1% improvement on MLP model and 39.3% on LSTM are achieved. Figure 10 presents the performance regression for both the *Raw N* = 1 and benchmark MLP models, along with the corresponding prediction errors. In the benchmark model, the prediction output exhibit fluctuations, this behavior is reflected in the higher RMSE of the model performance. However, with the *Raw _N* = 1 model, the variance of the error is notably reduced, indicating more stable predictions with less fluctuation.

### 3.1.3. Effect of K-means Clustering

Figure 11 presents the clustering results for the weather data, including dry temperature, humidity, and wind speed. While clear boundaries are observable for $K$ = 2 and $K$ = 3, the boundaries become less distinct when the number of clusters

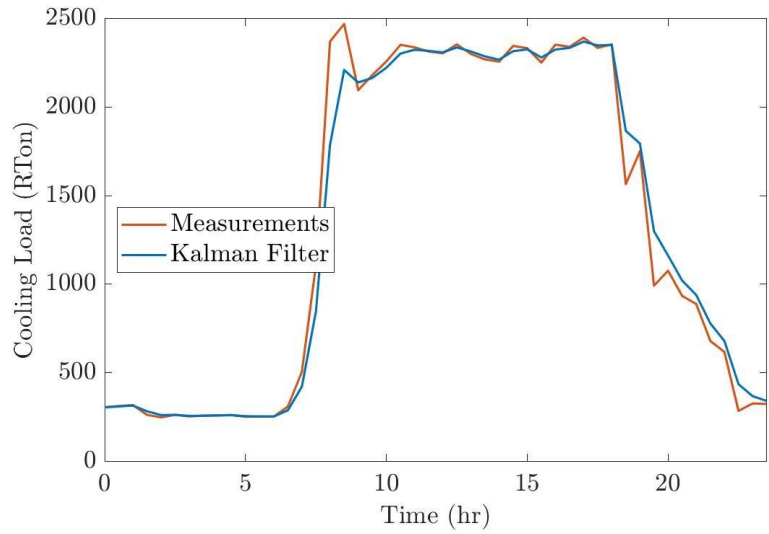

Figure 9: Cooling Load Measurement and Kalman Filter

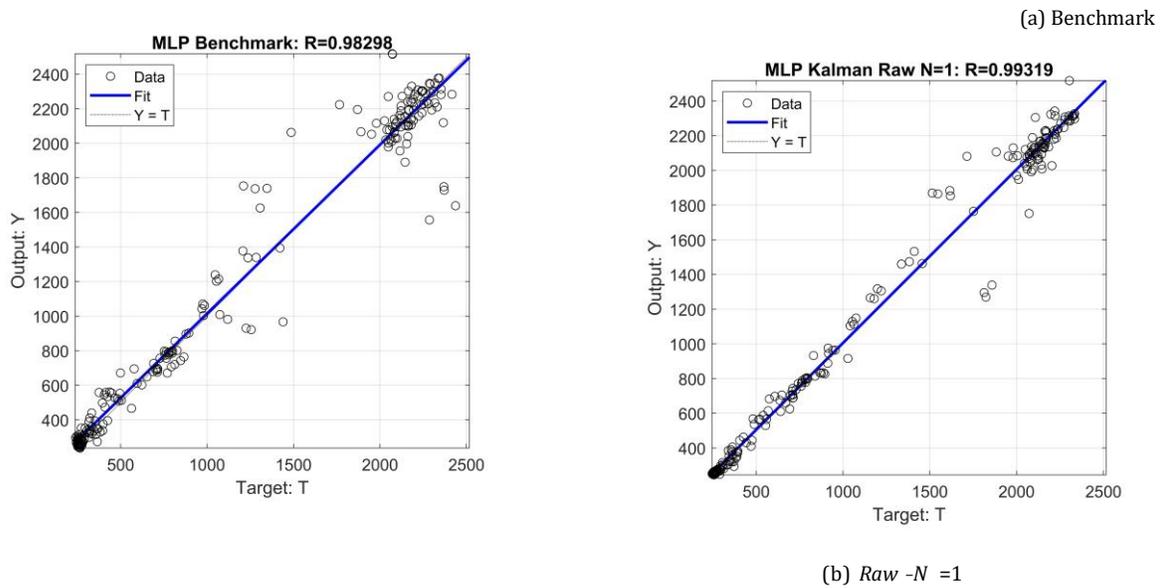

(a) Benchmark

(b) *Raw -N* =1

Figure 10: Performance Regression on Test Dataset for Benchmark and *Raw N* = 1 Feature Sets on MLP

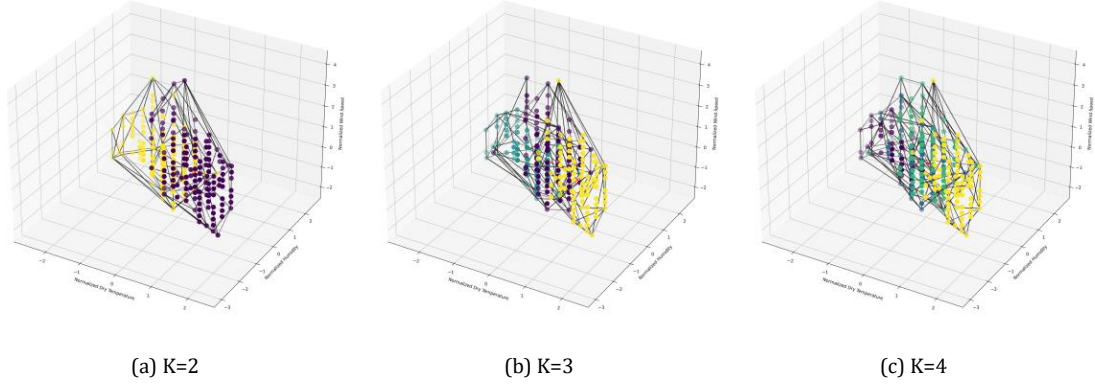

| (a) K=2 | (b) K=3 | (c) K=4 |

Figure 11: K-means Clustering on Dry Temperature, Humidity and Wind Speed

is set to 4. As a result, clustering for values of *K* greater than 4 was not investigated.

Comparing the RMSE for feature sets with different weather clusters, the *K* = 2 feature set outperforms all raw weather feature sets, indicating that weather clustering not only reduces model complexity but also enhances overall prediction performance. For *N* = 1 feature sets in both MLP and LSTM models, the RMSE increases as *K* values rise, suggesting that increasing the number of weather clusters increases the likelihood of model overfitting on the test dataset. Weather clustering on *N* = 5 feature sets does not necessarily improve test performance, which may be due to the fact that including more historical cooling load data makes the model more complex and prone to overfitting.

Overall, the *K* = 2 *N* = 1 feature set outperforms others on both MLP and LSTM models. For MLP, it achieves a 14.1% improvement compared to *Raw N* = 1 and a 46.5% improvement over the benchmark. For LSTM, it achieves a 4.5% improvement over *Raw N* = 1 and a 41.8% improvement over the benchmark. Notably, most *N* = 1 feature sets outperform all *N* = 5 feature sets, except for *Raw N* = 1 on MLP and *K* = 4 *N* = 1 on LSTM. This suggests that including more historical cooling load data does not necessarily improve prediction accuracy, as the current cooling load itself is a comprehensive feature that captures the necessary information for half-hour-ahead predictions. These results emphasize the importance of constructing balanced feature sets to prevent overfitting in load prediction models.

### 3.1.4. MLP vs. LSTM

The comparison results between MLP and LSTM models suggests that, despite being a simpler neural network, the performance of the MLP models can surpass that of the LSTM model for the purpose of this study with the best-performing MLP outperforms LSTM by 10.4% on *K* = 2 *N* = 1 feature set. It it also noted that MLP models generally can outperform LSTM models on feature sets with fewer parameters, while LSTM performs better than MLP on more complex feature sets with more parameters ($N$ = 5, $K$ = 3,4). This observation supports the argument that current cooling load is a more useful feature than previous time steps, as LSTM inherently captures that information. Additionally, the MLP model offers significant computational savings, reducing the training time from over a minute with LSTM to within 5 seconds.

### 3.2. Chiller Optimization
#### 3.2.1. Chiller Model

There are a total of six chillers installed at the building for this study, four of which are of 1000RT capacity and the other two are of 500RT capacity. As one number of 1000RT and 500RT chiller are served as standby chiller in case of malfunction of equipment, only four chillers are modelled into the optimisation model. The optimization model is hence for the three 1000RT chiller and one 500RT chiller.

Table 4: Chiller Part Load Performance Data

| PLR | 1000RT Chiller | | 500RT Chiller | |
|---|---|---|---|---|
| | Chiller Eff, kW/RT | Chiller power, kW | Chiller Eff, kW/RT | Chiller power, kW |
| 0.2 | 0.767 | 153 | 0.822 | 82 |
| 0.3 | 0.632 | 190 | 0.686 | 103 |
| 0.4 | 0.557 | 223 | 0.612 | 122 |
| 0.5 | 0.514 | 257 | 0.564 | 141 |
| 0.6 | 0.491 | 295 | 0.536 | 161 |
| 0.7 | 0.475 | 333 | 0.516 | 181 |
| 0.8 | 0.467 | 374 | 0.5 | 200 |
| 0.9 | 0.465 | 419 | 0.495 | 223 |
| 1 | 0.483 | 483 | 0.498 | 249 |

The chiller efficiency and chiller power at part load ratio of the installed 1000RT and 500RT chiller onsite is obtained from the equipment specifications as specified in Table 4. The mathematical model of the chiller part load power is represented by PLR for both chillers as explained in Section 2.4.

#### 3.2.2. Chiller Optimization

The simulation results indicate that the optimization algorithm can potentially achieve a 13.8% energy saving compared to the actual metered consumption, primarily through an optimized chiller sequencing strategy that facilitates a more efficient chiller combination during peak daytime loads. During typical weekdays, the building's peak load ranges from 2100RT to 2350RT and the current sequencing employs two 1000RT chillers and one 500RT chiller to support the peak load. The optimized strategy as detailed in Table 5 and 6 however suggests that operating three 1000RT chillers is more efficient than the current sequencing during peak load. This result can be intuitively understood by comparing the PLR efficiency of the 1000RT and 500RT chillers. The 1000RT chiller is more efficient at 60% load than the 500RT chiller at 90% load, making it more energy-efficient to run the 1000RT chillers above 60% load rather than maxing out the 500RT chiller. In practice, this suggested strategy can be implemented by adjusting chiller loading through controlling the chilled water flow rate and supply temperature to align with the PLR recommendations.

It's important to note that the potential saving is based on two assumptions. First, since the optimization is based on predicted Kalman filter loads, it is assumed that the Kalman filter effectively minimizes the noise present in raw measurements, making the filtered load a more accurate representation of the actual cooling load. The difference between the Kalman filter load and the predicted load is then neglected due to the high accuracy achieved by the best prediction model. Second, the chiller optimization relies on the chiller PLR performance curve provided by the manufacturer, which may represent higher efficiency than the actual operating conditions, as years of operation can degrade chiller performance. However, the results still serve as a reliable indicator of potential savings that could be achieved with necessary overhauls and proper maintenance to keep the chiller operating at optimal efficiency. Importantly, the proposed chiller sequencing strategy remains unchanged, even with chiller performance degradation, as any degradation is likely to affect chiller efficiency uniformly across various PLR levels.

Table 6: Optimized Chiller PLR with Total Power

| Time | Predicted Load | PLR | | | | Chiller Total Power (kW) |
|---|---|---|---|---|---|---|
| | | Chiller 1 | Chiller 2 | Chiller 3 | Chiller 4 | |
| 8:30  | 2267.2 | 0.86 | 0.63 | 0.78 | 0.00 | 1069.0 |
| 9:00  | 2233.7 | 0.76 | 0.76 | 0.71 | 0.00 | 1046.8 |
| 9:30  | 2181.4 | 0.72 | 0.68 | 0.78 | 0.00 | 1025.3 |
| 10:00 | 2283.2 | 0.75 | 0.73 | 0.81 | 0.00 | 1068.8 |
| 10:30 | 2291.3 | 0.80 | 0.73 | 0.76 | 0.00 | 1072.1 |
| 11:00 | 2326.0 | 0.76 | 0.78 | 0.78 | 0.00 | 1086.7 |
| 11:30 | 2348.9 | 0.79 | 0.78 | 0.78 | 0.00 | 1096.4 |
| 12:00 | 2323.4 | 0.76 | 0.76 | 0.80 | 0.00 | 1085.7 |
| 12:30 | 2335.7 | 0.80 | 0.74 | 0.80 | 0.00 | 1091.3 |
| 13:00 | 2324.9 | 0.76 | 0.78 | 0.78 | 0.00 | 1085.8 |
| 13:30 | 2297.9 | 0.76 | 0.79 | 0.75 | 0.00 | 1074.5 |
| 14:00 | 2327.2 | 0.83 | 0.77 | 0.73 | 0.00 | 1088.6 |
| 14:30 | 2291.2 | 0.80 | 0.73 | 0.76 | 0.00 | 1072.1 |
| 15:00 | 2284.2 | 0.73 | 0.79 | 0.77 | 0.00 | 1068.7 |
| 15:30 | 2293.1 | 0.73 | 0.80 | 0.76 | 0.00 | 1072.8 |
| 16:00 | 2219.3 | 0.74 | 0.74 | 0.74 | 0.00 | 1040.5 |
| 16:30 | 2188.4 | 0.70 | 0.80 | 0.69 | 0.00 | 1029.4 |
| 17:00 | 2148.3 | 0.75 | 0.78 | 0.61 | 0.00 | 1015.1 |
| 17:30 | 2135.3 | 0.77 | 0.66 | 0.71 | 0.00 | 1007.3 |
| 18:00 | 2109.3 | 0.73 | 0.69 | 0.69 | 0.00 | 995.6  |

### 3.3. Thermal Energy Storage
### 3.3.1. Design Proposals

Thermal energy storage (TES) tanks offer a means to reduce energy costs by storing excess chilled water during off-peak hours when electrical tariffs are lower. However, their adoption in buildings, particularly in land-scarce countries like

Singapore, is limited due to increased upfront capital costs and significant space requirements. This section compares the existing chiller plant design with configurations incorporating TES tanks, evaluating each option from the perspectives of energy consumption and cost. The optimization model is used to ensure maximum chiller efficiency across all chiller-TES configurations.

The comparison baseline is the current chiller plant installation with optimized chiller sequencing as detailed in Section 3.2. The chiller-TES design proposals are based on the predicted cooling load profile from Section 3.1.1, with parameters listed in Table 7. Four configurations are studied, varying in chiller and TES capacity and the details illustrated in Table 8.

Table 7: Input parameters for Chiller-TES Model design analysis

| Parameter | Value | Unit |
| --- | --- | --- |
| Peak Electricity Tariff | 0.2967 | $/kWh |
| Off-peak Electricity Tariff | 0.1843 | $/kWh |
| Contract Capacity Charge | 16.48 | $/kW |
| Capital Cost for Chiller [73] | 654.00 | $/kW |
| Capital Cost for TES [73] | 71.09 | $/kWh |

Table 8: Comparison of Chiller and TES specifications across different proposals

| Item | Unit | Baseline | Proposal 1 | Proposal 2 | Proposal 3 | Proposal 4 |
| --- | --- | --- | --- | --- | --- | --- |
| **Chiller** | | | | | | |
| 1000RT Chiller | no. | 3 | 2 | 2 | 3 | 1 |
| 500RT Chiller | no. | 1 | 0 | 1 | 1 | 1 |
| Chiller Capacity | kW | 12309.5 | 7034.0 | 8792.5 | 12309.5 | 5275.5 |
| **TES** | | | | | | |
| TES Charging | kWh | 0.0 | 20217.4 | 51879.2 | 74394.1 | 35809.3 |
| TES Discharging | kWh | 0.0 | 20202.2 | 51860.5 | 74347.3 | 35776.5 |
| TES Capacity | kWh | 0.0 | 20217.4 | 51879.2 | 74394.1 | 35809.3 |
| Percentage to Total Daily Cooling Load | | 0% | 20% | 48% | 69% | 33% |

Chiller sequencing strategy is adjusted to adapt to the different design configurations to maximize chiller operating efficiency of each proposal. In Proposal 1, a single 1000RT chiller operates at maximum efficiency loading during off-peak hours to support the night load and charge the TES, with two 1000RT chillers handling day load with TES discharge. In proposal 2, three chillers run at optimized PLR during off-peak hours to charge the TES, with one 1000RT chiller running at maximum efficiency loading during peak hours with TES discharge. Proposal 3 also have three 1000RT chillers operating at off-peak hours for charging of TES, with one 500RT chiller operating at maximum efficiency loading during peak hours. Lastly, Proposal 4 maintains two chillers at an evenly loading at optimized efficiency loading throughout the day and make use of TES to balance load charging and discharging during off-peak and peak hour.

The operations of the four proposed Chiller-TES models are illustrated from Figure 12 to 15, where various elements are visually differentiated for clarity. The blue lines depict the cooling load as forecasted by the prediction model. Orange blocks indicate the portion of the load that the chillers are actively supplying to the building. Grey blocks signify the surplus load by chiller that is not immediately used and is instead stored in the TES system. Lastly, green blocks illustrate the periods during which the TES discharges, making up to the building's cooling load. This color-coded representation facilitates an intuitive understanding of how the chiller and TES systems interact and operate in different configurations.

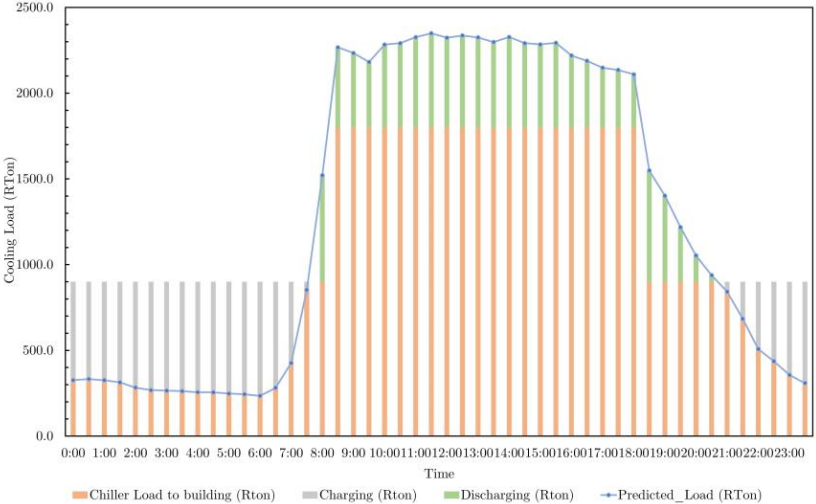

Figure 12: Chiller-TES Proposal 1

*3.3.2. Comparative Analysis*

A comparative analysis, detailed in Table 9 evaluates the energy consumption and cost of various chiller-TES configurations. Proposals 1 and 4 stand out for the reduced capital costs, with savings of 25% and 26% respectively, attributes to the deployment of fewer chillers and smaller TES units. Proposal 3 offers the highest operational cost savings at 19%, taking advantage of the differential between peak and off-peak electrical tariffs and its large TES capacity. However, this comes at the expense of a substantial increase in capital investment that is 66% higher than the benchmark. Proposal 4 emerges as the most balanced option that can deliver capital and operational cost savings of 26% and 13% respectively. It also has the best long-term cost savings of 17% over a decade.

These findings suggest a viable alternative for designing chiller plants in commercial settings, such as office buildings and shopping centers, where there is a considerable difference between peak and non-peak load demands (85% difference noted in this study). Traditional setups typically require more chillers to meet peak demands, whereas non-peak periods are managed with one unit. The strategic integration of TES within the chiller plant can yield substantial savings in both capital and operating costs. These savings come from: (a) lesser chiller installation, (b) tariff difference between peak and off-peak hours, and (c) reduction in maximum power demand, hence a lesser contracted capacity charge. Of the four TES models evaluated, the fourth proposal stands out, offering the most significant cost benefits in both capital and operational expenditures. The design principle of this proposal is to distribute the total daily load evenly and operate the same chillers at consistent loading throughout the day, thus maximizing the benefits in contracted capacity charge reduction while also leveraging the advantages of lesser chiller installation and the electrical tariff difference during peak and off-peak hours.

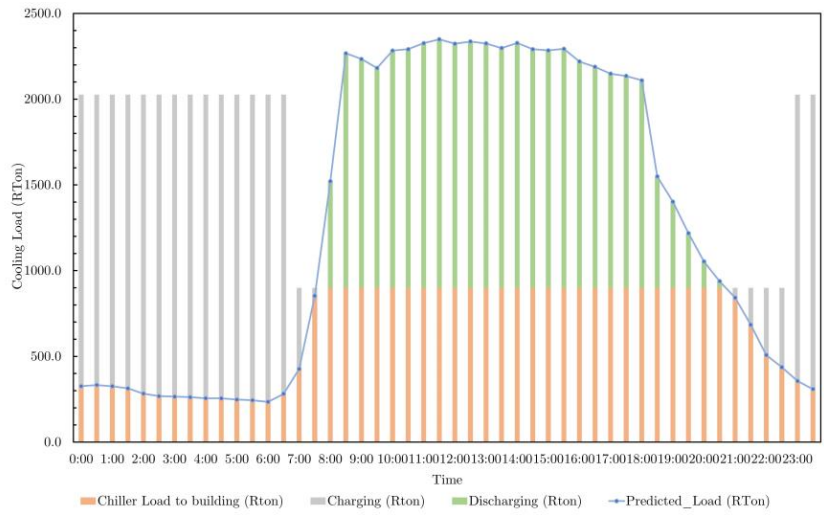

Figure 13: Chiller-TES Proposal 2

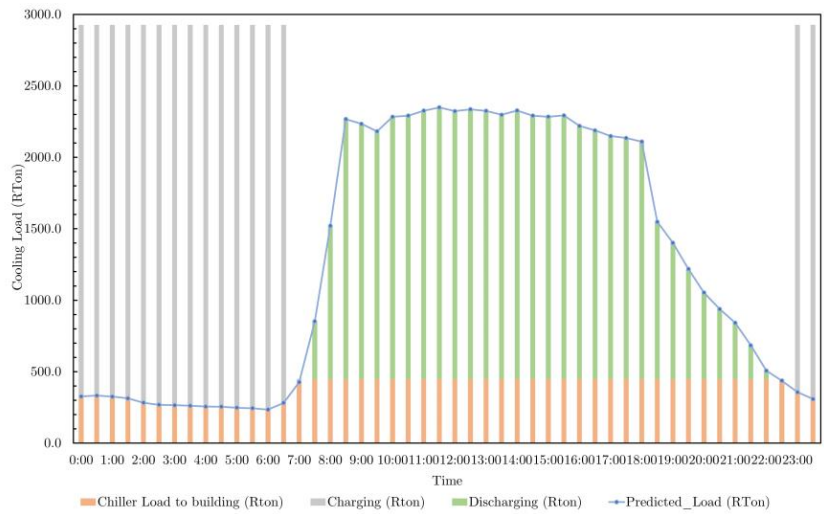

Figure 14: Chiller-TES Proposal 3

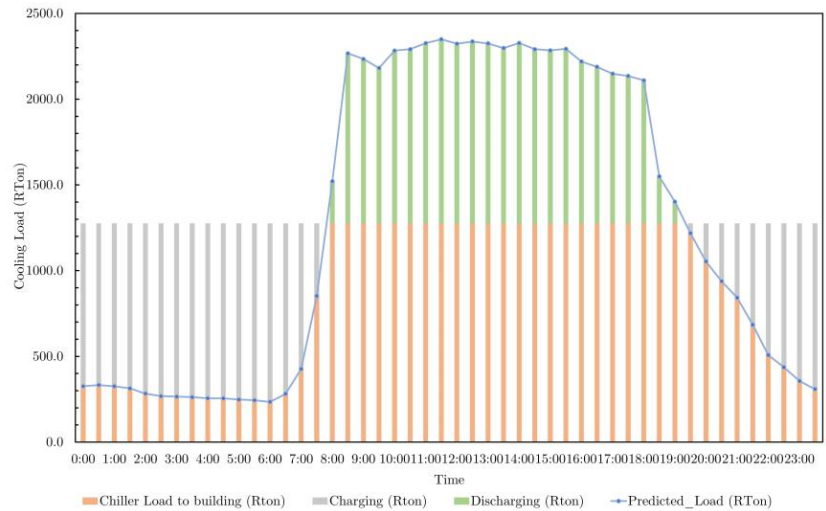

Figure 15: Chiller-TES Proposal 4

Table 9: Energy and Cost Analysis of Benchmark Model and TES Proposals

| Item | Unit | Baseline | Proposal 1 | Proposal 2 | Proposal 3 | Proposal 4 |
|---|---|---|---|---|---|---|
| **Energy Analysis** | | | | | | |
| Total Power Consumption | kWh | 14470.6 | 14232.3 | 14368.5 | 14600.8 | 14603.5 |
| Peak Power Consumption | kWh | 13295.7 | 10882.4 | 6696.0 | 3564.0 | 9735.7 |
| Off-peak Power Consumption | kWh | 1174.9 | 3349.9 | 7672.5 | 11036.8 | 4867.8 |
| Maximum Power Demand | kW | 1096.4 | 837.0 | 959.1 | 1379.6 | 608.5 |
| **Cost Analysis** | | | | | | |
| Chiller Capital Cost | $ | 8050413.0 | 4600236.0 | 5750295.0 | 8050413.0 | 3450177.0 |
| TES Capital Cost | $ | N/A | 1437256.1 | 3688092.9 | 5288676.6 | 2545682.1 |
| Total Capital Cost | $ | 8050413.0 | 6037492.1 | 9438387.9 | 13339089.6 | 5995859.1 |
| Percentage Savings in Capital Cost | | N/A | 25% | -17% | -66% | 26% |
| Electricity Tariff Cost in Peak Hours | $/day | 3944.8 | 3228.8 | 1986.7 | 1057.4 | 2888.6 |
| Electricity Tariff Cost in Off-peak Hours | $/day | 216.5 | 617.4 | 1414.0 | 2034.1 | 897.1 |
| Total Electricity Tariff Cost | $/day | 4161.4 | 3846.2 | 3400.7 | 3091.5 | 3785.7 |
| Contract Capacity Cost per Month | $/month | 18069.3 | 13793.8 | 15806.0 | 22735.8 | 10027.8 |
| Total Yearly Operating Cost | $ | 1735730.7 | 1569382.9 | 1430942.0 | 1401234.9 | 1502120.5 |
| Percentage Savings in Operating Cost | | N/A | 10% | 18% | 19% | 13% |
| 10 Years Costing | $ | 25407720.09 | 21731321.41 | 23747808.32 | 27351438.39 | 21017064.07 |
| Percentage Savings over 10 years | | N/A | 14% | 7% | -8% | 17% |

## 4. Conclusion

In this study, a comprehensive model predictive control strategy was developed for optimizing chiller plant performance of a commercial building in Singapore central business district, comprising load prediction, chiller sequencing optimization and integration of TES. Specifically, this study made the following contributions to the field of load prediction:

1. This study demonstrated the Kalman Filter's efficacy in refining raw cooling load data, which is inherently noisy due to chiller startup operations and measurement inaccuracies. The application of the Kalman Filter clarified the actual cooling load, enhancing the performance of both MLP and LSTM models by 23.1% and 39.3%, respectively.

2. The integration of K-means clustering to classify raw weather data led to a substantial 30.5% improvement in prediction accuracy when the cluster is set to 2. This demonstrates that clustering weather data can effectively reduce model complexity and enhance prediction accuracy.

Utilizing the results from load prediction, genetic algorithm optimization identified the most efficient chiller sequencing strategy based on optimal part-load performance rather than the conventional on/off operation. This optimized sequencing can potentially yield a notable 13.8% improvement in energy savings without incurring extensive equipment upgrades. Furthermore, while conventional chiller plant designs often neglect thermal energy storage due to upfront costs and space limitation, this study highlighted that with appropriate sizing guided by prediction and optimization results, chiller-TES design can reduce the numbers of chiller installation required and potentially offer significant reductions in both capital and operational expenses. In conclusion, the strategies outlined in this study provide a holistic approach in optimizing chiller plant from design to operation, provides actionable insights for building owners and design consultancy firms to streamline chiller plant performance effectively.

## References

[1] L. P´erez-Lombard, J. Ortiz, C. Pout, A review on buildings energy consumption information, Energy and Buildings 40 (2008) 394–398. doi:`10.1016/j.enbuild. 2007.03.007.`

[2] K. J. Chua, S. K. Chou, W. M. Yang, J. Yan, Achieving better energy-e cient air conditioning – A review of technologies and strategies, Applied Energy 104 (2013) 87–104. doi:`10.1016/j.apenergy.2012.10.037.`

[3] L. Jayamaha, Energy-Efficient Building Systems: Green Strategies for Operation and Maintenance, McGraw Hill Professional, 2006.

[4] S. Wang, Z. Ma, Supervisory and Optimal Control of Building HVAC Systems: A Review, HVAC&R Research 14 (2011) 3–32. doi:`10.1080/10789669.2008.10390991`.

[5] K. He, Q. Fu, Y. Lu, Y. Wang, J. Luo, H. Wu, J. Chen, Predictive control optimization of chiller plants based on deep reinforcement learning, Journal of Building Engineering 76 (2023). doi:`10.1016/j.jobe.2023.107158`.

[6] Y. Yao, D. K. Shekhar, State of the art review on model predictive control (MPC) in Heating Ventilation and Air-conditioning (HVAC) field, Building and Environment 200 (2021). doi:`10.1016/j.buildenv.2021.107952`.

[7] P. H. Shaikh, N. B. M. Nor, P. Nallagownden, I. Elamvazuthi, T. Ibrahim, A review on optimized control systems for building energy and comfort management of smart sustainable buildings, Renewable and Sustainable Energy Reviews 34 (2014) 409–429. doi:`10.1016/j.rser.2014.03.027`.

[8] C. Fan, F. Xiao, Y. Zhao, A short-term building cooling load prediction method using deep learning algorithms, Applied Energy 195 (2017) 222–233. doi:`10.1016/j.apenergy.2017.03.064`.

[9] J. F. Kreider, D. E. Claridge, P. Curtiss, R. Dodier, J. S. Haberl, M. Krarti, Building Energy Use Prediction and System Identification Using Recurrent Neural Networks, Journal of Solar Energy Engineering 117 (1995) 161–166. doi:`10.1115/1.2847757`.

[10] J. F. Kreider, Improved artificial neural networks for commercial building energy prediction, in: ASME-JSES-KSES International Solar Energy Conference, volume 1, 1992, pp. 361–366.

[11] P. S. Curtiss, J. F. Kreider, M. J. Brandemuehl, Artificial neural networks proof of concept for local and global control of commercial building HVAC systems., ASME, NEW YORK, NY(USA). (1993) 429–443.

[12] S. A. Kalogirou, M. Bojic, Artificial neural networks for the prediction of the energy consumption of a passive solar building (2000).

[13] A. H. Neto, F. A. S. Fiorelli, Comparison between detailed model simulation and artificial neural network for forecasting building energy consumption, Energy and Buildings 40 (2008) 2169–2176. doi:`10.1016/j.enbuild.2008.06.013`.

[14] S. S. K. Kwok, E. W. M. Lee, A study of the importance of occupancy to building cooling load in prediction by intelligent approach, Energy Conversion and Management 52 (2011) 2555–2564. doi:`10.1016/j.enconman.2011.02.002`.

[15] M. Yalcintas, S. Akkurt, Artificial neural networks applications in building energy predictions and a case study for tropical climates, International Journal of Energy Research 29 (2005) 891–901. doi:`10.1002/er.1105`.

[16] B. Bektas Ekici, U. T. Aksoy, Prediction of building energy needs in early stage of design by using ANFIS, Expert Systems with Applications 38 (2011) 5352–5358. doi:`10.1016/j.eswa.2010.10.021`.

[17] S. Sholahudin, H. Han, Simplified dynamic neural network model to predict heating load of a building using Taguchi method, Energy 115 (2016) 1672–1678. doi:`10.1016/j.energy.2016.03.057`.

[18] E. Sala-Cardoso, M. Delgado-Prieto, K. Kampouropoulos, L. Romeral, Activityaware HVAC power demand forecasting, Energy and Buildings 170 (2018) 15–24. doi:`10.1016/j.enbuild.2018.03.087`.

[19] M. W. Ahmad, M. Mourshed, Y. Rezgui, Trees vs Neurons: Comparison between random forest and ANN for high-resolution prediction of building energy consumption, Energy and Buildings 147 (2017) 77–89. doi:`10.1016/j.enbuild.2017.04.038`.

[20] S.-M. Hong, G. Paterson, D. Mumovic, P. Steadman, Improved benchmarking comparability for energy consumption in schools, Building Research & Information 42 (2014) 47–61. doi:`10.1080/09613218.2013.814746`.

[21] A. Kavousian, R. Rajagopal, Data-Driven Benchmarking of Building Energy Efficiency Utilizing Statistical Frontier Models, Journal of Computing in Civil Engineering 28 (2014) 79–88. doi:`10.1061/(ASCE)CP.1943-5487.0000327`.

[22] Y. Zhang, Z. O'Neill, B. Dong, G. Augenbroe, Comparisons of inverse modeling approaches for predicting building energy performance, Building and Environment 86 (2015) 177–190. doi:`10.1016/j.buildenv.2014.12.023`.

[23] W. Chung, Review of building energy-use performance benchmarking methodologies, Applied Energy 88 (2011) 1470–1479. doi:`10.1016/j.apenergy.2010.`

11.022.

[24] R. E. Edwards, J. New, L. E. Parker, Predicting future hourly residential electrical consumption: A machine learning case study, Energy and Buildings 49 (2012) 591–603. doi:`10.1016/j.enbuild.2012.03.010`.

[25] J. Massana, C. Pous, L. Burgas, J. Melendez, J. Colomer, Short-term load forecasting in a non-residential building contrasting models and attributes, Energy and Buildings 92 (2015) 322–330. doi:`10.1016/j.enbuild.2015.02.007`.

[26] C. Fan, Y. Ding, Cooling load prediction and optimal operation of HVAC systems using a multiple nonlinear regression model, Energy and Buildings 197 (2019) 7–17. doi:`10.1016/j.enbuild.2019.05.043`.

[27] A. Rahman, V. Srikumar, A. D. Smith, Predicting electricity consumption for commercial and residential buildings using deep recurrent neural networks, Applied Energy 212 (2018) 372–385. doi:`10.1016/j.apenergy.2017.12.051`.

[28] N. Tarkhan, J. T. Szcześniak, C. Reinhart, Façade feature extraction for urban performance assessments: Evaluating algorithm applicability across diverse building morphologies, Sustainable Cities and Society 105 (2024) 105280. doi:`10. 1016/j.scs.2024.105280`.

[29] S. Lawrence, C. Giles, Overfitting and neural networks: Conjugate gradient and backpropagation, in: Proceedings of the IEEE-INNS-ENNS International Joint Conference on Neural Networks. IJCNN 2000. Neural Computing: New Challenges and Perspectives for the New Millennium, IEEE, Como, Italy, 2000.

[30] M. Dahl, A. Brun, G. B. Andresen, Using ensemble weather predictions in district heating operation and load forecasting, Applied Energy 193 (2017) 455–465. doi:`10.1016/j.apenergy.2017.02.066`.

[31] H. S. Lim, G. Kim, Prediction model of Cooling Load considering time-lag for preemptive action in buildings, Energy and Buildings 151 (2017) 53–65. doi:`10.1016/j.enbuild.2017.06.019`.

[32] C. Zhang, Y. Zhao, X. Zhang, C. Fan, T. Li, An Improved Cooling Load Prediction Method for Buildings with the Estimation of Prediction Intervals, Procedia Engineering 205 (2017) 2422–2428. doi:`10.1016/j.proeng.2017.09. 967`.

[33] Z. Zheng, Z. Zhuang, Z. Lian, Y. Yu, Study on Building Energy Load Prediction Based on Monitoring Data, Procedia Engineering 205 (2017) 716–723. doi:10.1016/j.proeng.2017.09.894.

[34] Q. Li, Q. Meng, J. Cai, H. Yoshino, A. Mochida, Predicting hourly cooling load in the building: A comparison of support vector machine and different artificial neural networks, Energy Conversion and Management 50 (2009) 90–96.

[35] Y. Guo, E. Nazarian, J. Ko, K. Rajurkar, Hourly cooling load forecasting using time-indexed ARX models with two-stage weighted least squares regression, Energy Conversion and Management 80 (2014) 46–53.

[36] F. Tang, A. Kusiak, X. Wei, Modeling and short-term prediction of HVAC system with a clustering algorithm, Energy and Buildings 82 (2014) 310–321. doi:10.1016/j.enbuild.2014.07.037.

[37] F. Aqlan, A. Ahmed, K. Srihari, M. T. Khasawneh, Integrating artificial neural networks and cluster analysis to assess energy efficiency of buildings, in: IIE Annual Conference. Proceedings, Institute of Industrial and Systems Engineers (IISE), 2014, p. 3936.

[38] J. G. Jetcheva, M. Majidpour, W.-P. Chen, Neural network model ensembles for building-level electricity load forecasts, Energy and Buildings 84 (2014) 214–223. doi:10.1016/j.enbuild.2014.08.004.

[39] P. Duan, K. Xie, T. Guo, X. Huang, Short-Term Load Forecasting for Electric Power Systems Using the PSO-SVR and FCM Clustering Techniques, Energies 4 (2011) 173–184. doi:10.3390/en4010173.

[40] L. Xuemei, D. Yuyan, D. Lixing, J. Liangzhong, Building cooling load forecasting using fuzzy support vector machine and fuzzy C-mean clustering, in: 2010 International Conference on Computer and Communication Technologies in Agriculture Engineering, volume 1, 2010, pp. 438–441. doi:10.1109/CCTAE.2010.5543577.

[41] J. Yang, C. Ning, C. Deb, F. Zhang, D. Cheong, S. E. Lee, C. Sekhar, K. W. Tham, K-Shape clustering algorithm for building energy usage patterns analysis and forecasting model accuracy improvement, Energy and Buildings 146 (2017) 27–37. doi:10.1016/j.enbuild.2017.03.071.

[42] C. Fan, F. Xiao, S. Wang, Development of prediction models for next-day building energy consumption and peak power demand using data mining techniques, Applied Energy 127 (2014) 1–10. doi:10.1016/j.apenergy.2014.04.016.

[43] M. Santamouris, G. Mihalakakou, P. Patargias, N. Gaitani, K. Sfakianaki, M. Papaglastra, C. Pavlou, P. Doukas, E. Primikiri, V. Geros, M. N. Assimakopoulos, R. Mitoula, S. Zerefos, Using intelligent clustering techniques to classify the energy performance of school buildings, Energy and Buildings 39 (2007) 45–51. doi:`10.1016/j.enbuild.2006.04.018`.

[44] Z. Yu, B. C. M. Fung, F. Haghighat, H. Yoshino, E. Morofsky, A systematic procedure to study the influence of occupant behavior on building energy consumption, Energy and Buildings 43 (2011) 1409–1417. doi:`10.1016/j.enbuild.2011.02.002`.

[45] F. Xiao, C. Fan, Data mining in building automation system for improving building operational performance, Energy and Buildings 75 (2014) 109–118. doi:`10.1016/j.enbuild.2014.02.005`.

[46] F. McLoughlin, A. Duffy, M. Conlon, A clustering approach to domestic electricity load profile characterisation using smart metering data, Applied Energy 141 (2015) 190–199. doi:`10.1016/j.apenergy.2014.12.039`.

[47] L. Zhang, J. Wen, Y. Li, J. Chen, Y. Ye, Y. Fu, W. Livingood, A review of machine learning in building load prediction, Applied Energy 285 (2021) 116452. doi:`10.1016/j.apenergy.2021.116452`.

[48] M. A. Asghar, M. J. Khan, M. Rizwan, R. M. Mehmood, S.-H. Kim, An Innovative Multi-Model Neural Network Approach for Feature Selection in Emotion Recognition Using Deep Feature Clustering, Sensors 20 (2020) 3765. doi:`10.3390/s20133765`.

[49] J. J. Shi, Clustering Technique for Evaluating and Validating Neural Network Performance, Journal of Computing in Civil Engineering 16 (2002) 152–155. doi:`10.1061/(ASCE)0887-3801(2002)16:2(152)`.

[50] D. Fay, J. V. Ringwood, On the Influence of Weather Forecast Errors in ShortTerm Load Forecasting Models, IEEE Transactions on Power Systems 25 (2010) 1751–1758. doi:`10.1109/TPWRS.2009.2038704`.

[51] W.-H. Chen, F. You, Sustainable energy management and control for Decarbonization of complex multi-zone buildings with renewable solar and geothermal energies using machine learning, robust optimization, and predictive control, Applied Energy 372 (2024) 123802. doi:`10.1016/j.apenergy.2024.123802`.

[52] F. Tang, A. Kusiak, X. Wei, Modeling and short-term prediction of HVAC system with a clustering algorithm, Energy and Buildings 82 (2014) 310–321. doi:`10.1016/j.enbuild.2014.07.037`.

[53] J. G. Jetcheva, M. Majidpour, W.-P. Chen, Neural network model ensembles for building-level electricity load forecasts, Energy and Buildings 84 (2014) 214–223. doi:`10.1016/j.enbuild.2014.08.004`.

[54] F. Aqlan, A. Ahmed, K. Srihari, M. T. Khasawneh, Integrating artificial neural networks and cluster analysis to assess energy efficiency of buildings, in: IIE Annual Conference. Proceedings, Institute of Industrial and Systems Engineers (IISE), 2014, p. 3936.

[55] Z. Yu, B. C. M. Fung, F. Haghighat, H. Yoshino, E. Morofsky, A systematic procedure to study the influence of occupant behavior on building energy consumption, Energy and Buildings 43 (2011) 1409–1417. doi:`10.1016/j.enbuild.2011.02.002`.

[56] W. Xu, B. Svetozarevic, L. Di Natale, P. Heer, C. N. Jones, Data-driven adaptive building thermal controller tuning with constraints: A primal–dual contextual Bayesian optimization approach, Applied Energy 358 (2024) 122493. doi:`10.1016/j.apenergy.2023.122493`.

[57] J. Drgoňa, A. R. Tuor, V. Chandan, D. L. Vrabie, Physics-constrained deep learning of multi-zone building thermal dynamics, Energy and Buildings 243 (2021) 110992. doi:`10.1016/j.enbuild.2021.110992`.

[58] K. C. Chan, V. T. T. Wong, A. K. F. Yow, P. L. Yuen, C. Y. H. Chao, Development and performance evaluation of a chiller plant predictive operational control strategy by artificial intelligence, Energy and Buildings 262 (2022). doi:`10.1016/j.enbuild.2022.112017`.

[59] C. Zhang, J. Li, Y. Zhao, T. Li, Q. Chen, X. Zhang, A hybrid deep learningbased method for short-term building energy load prediction combined with an interpretation process, Energy and Buildings 225 (2020). doi:`10.1016/j.enbuild.2020.110301`.

[60] L. Wang, E. W. M. Lee, R. K. K. Yuen, Novel dynamic forecasting model for building cooling loads combining an artificial neural network and an ensemble approach, Applied Energy 228 (2018) 1740–1753. doi:`10.1016/j.apenergy.2018.07.085`.

[61] C. Zhuang, R. Choudhary, A. Mavrogianni, Probabilistic occupancy forecasting for risk-aware optimal ventilation through autoencoder Bayesian deep neural networks, Build Environ 219 (2022) 109207. doi:`10.1016/j.buildenv.2022. 109207`.

[62] Z. Wang, T. Hong, M. A. Piette, Building thermal load prediction through shallow machine learning and deep learning, Applied Energy 263 (2020). doi:`10. 1016/j.apenergy.2020.114683`.

[63] C. Zhang, Z. Luo, Y. Rezgui, T. Zhao, Enhancing multi-scenario data-driven energy consumption prediction in campus buildings by selecting appropriate inputs and improving algorithms with attention mechanisms, Energy and Buildings 311 (2024). doi:`10.1016/j.enbuild.2024.114133`.

[64] W. Li, G. Gong, H. Fan, P. Peng, L. Chun, Meta-learning strategy based on user preferences and a machine recommendation system for real-time cooling load and COP forecasting, Applied Energy 270 (2020). doi:`10.1016/j.apenergy. 2020.115144`.

[65] Y. Jin, D. Yan, X. Kang, A. Chong, H. Sun, S. Zhan, Forecasting building occupancy: A temporal-sequential analysis and machine learning integrated approach, Energy and Buildings 252 (2021). doi:`10.1016/j.enbuild.2021. 111362`.

[66] X. Kang, X. Wang, J. An, D. Yan, A novel approach of day-ahead cooling load prediction and optimal control for ice-based thermal energy storage (TES) system in commercial buildings, Energy and Buildings 275 (2022). doi:`10.1016/ j.enbuild.2022.112478`.

[67] L. Zhang, J. Wen, A systematic feature selection procedure for short-term datadriven building energy forecasting model development, Energy and Buildings 183 (2019) 428–442. doi:`10.1016/j.enbuild.2018.11.010`.

[68] X. Li, J. Wen, Building energy consumption on-line forecasting using physics based system identification, Energy and Buildings 82 (2014) 1–12. doi:`10.1016/ j.enbuild.2014.07.021`.

[69] H. Dagdougui, F. Bagheri, H. Le, L. Dessaint, Neural network model for shortterm and very-short-term load forecasting in district buildings, Energy and Buildings 203 (2019) 109408. doi:`10.1016/j.enbuild.2019.109408`.

[70] J. Yu, Q. Liu, A. Zhao, X. Qian, R. Zhang, Optimal chiller loading in HVAC System Using a Novel Algorithm Based on the distributed framework, Journal of Building Engineering 28 (2020) 101044. doi:`10.1016/j.jobe.2019.101044`.

[71] W.-S. Lee, Y.-T. Chen, Y. Kao, Optimal chiller loading by differential evolution algorithm for reducing energy consumption, Energy and Buildings 43 (2011) 599–604. doi:`10.1016/j.enbuild.2010.10.028`.

[72] A. J. Ardakani, F. F. Ardakani, S. Hosseinian, A novel approach for optimal chiller loading using particle swarm optimization, Energy and Buildings 40 (2008) 2177–2187. doi:`10.1016/j.enbuild.2008.06.010`.

[73] K. Zaw, Z. Z. S. Kwik, W. Q. Chang, M. R. Islam, T. K. Poh, A technocommercial decision support framework for optimal district cooling system design in tropical regions, Applied Thermal Engineering 220 (2023). doi:`10.1016/j.applthermaleng.2022.119668`.